\documentclass[a4paper,11pt]{article}
\usepackage{silence}

\usepackage{feynmp-auto}
\DeclareMathAlphabet{\mathbbold}{U}{bbold}{m}{n}

\usepackage{dsfont,bbm}
\usepackage{graphicx}
\usepackage{fullpage}
\usepackage{framed}
\usepackage{caption,subcaption}

\usepackage{mystyle}
\graphicspath{{./figures/}}
\usepackage[table]{xcolor}
\usepackage{booktabs}
\usepackage{multirow}
\usepackage{pdflscape}

\usepackage{tikz}
\usetikzlibrary{decorations.pathmorphing, decorations.markings, calc, fadings,
decorations.pathreplacing, patterns,shapes.arrows,positioning,3d,arrows,decorations.markings,shadings,backgrounds}
\usepackage{pgfplots}

\usepackage[displaymath, mathlines]{lineno}
\usepackage{tensor}
\usepackage{cancel}
\usepackage{amsmath}

\usepackage{framed}
\usepackage{url}
\usepackage{caption}
\usepackage{silence}
\title{\sffamily  Boosted Higgs-pair production associated with large $E_T^{miss}$: a signal of $Z^\prime$}
\author{Chuan-Ren Chen, Yu-Xiang Lin, Hsien-Chih Wu and  Jason Yue}
\affiliation[]{
Department of Physics, National Taiwan Normal University, Taipei 116, Taiwan\\
}

\emailAdd{crchen@ntnu.edu.tw}
\emailAdd{80241002s@gapps.ntnu.edu.tw}
\emailAdd{60241026s@gapps.ntnu.edu.tw}
\emailAdd{jason.yue@ntnu.edu.tw}

\abstract{Non-resonant production of Higgs-pair via heavy intermediate states may be
  a distinctive signature for extended discrete symmetries when accompaied by large missing transvers energy. We discuss $T$-parity as an example of such symmetry within the Littlest Higgs Model, where a new heavy gauge boson $Z'$, the $T$-odd partner of SM $Z$-boson, predominately decays into to a Higgs boson and a dark matter candidate $\chi$. In essence, $T$-parity  stablises simultaneously both the Higgs mass and the dark matter. 
Production via  $pp\rightarrow Z' Z'\rightarrow 2h 2\chi$  may therefore yield important clues about symmetries connecting the Higgs and dark sectors.   This paper makes a case for  the search for this channel at the LHC by studying its discovery potential. It is demonstrated that in situations where a large  $Z'-\chi$ mass gap results in a boosted topology, the jet-substructure technique  can be leveraged to reach the required significance for discovery in the  $2h\rightarrow 2\gamma2b$ decay mode.   
    }

\begin{document}
\maketitle


\section{Introduction}

The success of the Large Hadron Collider (LHC) owes to the discovery of the last missing piece of the Standard Model (SM), the Higgs boson. 
Since then, it has been a  priority  to pin down the nature of electroweak symmetry breaking by measuring the Higgs couplings with the other SM particles. 
The parameters within the SM are completely determined by the observed the Higgs mass, yet no substantial deviations
have been observed in any of the measured couplings. 
On the other hand, observation of neutrino masses and oscillation, dark matter and the baryon asymmetry in the universe indicates that the SM is incomplete, thereby motivating for extensions.

Due to the discovery being made in the diboson decay modes, there is firm evidence that the  Higgs mechanism indeed operates in the electroweak gauge sector. On the other hand, significant progress in direct measurements of  Yukawa couplings has only been made in third generation, since the SM predicts that the fermion masses endowed by the Higgs mechanism should be proportional to the coupling strength.
Currently,
only $h\tau\tau$ coupling have been observed at $>5\sigma$  significance \cite{Khachatryan:2016vau}.
The gauge and fermionic couplings to the Higgs boson can be inferred from single Higgs boson production via current LHC data. On the other hand, Higgs self couplings are unlitmately required to reconstruct the Higgs potential, but remains relatively unconstrained.
%
%
At the LHC, ATLAS \cite{ATLAS:2016ixk,ATLAS:2016qmt,TheATLAScollaboration:2016ibb,ATL-PHYS-PUB-2017-001,ATL-PHYS-PUB-2016-024,Aad:2015xja,Aaboud:2016xco,Aad:2015uka,Aad:2014yja} and CMS   \cite{Sirunyan:2017djm,CMS:2017ihs,CMS:2017ums,CMS:2016cdj,CMS:2016knm,CMS:2016ymn,CMS:2016tlj,CMS:2016ugf,CMS:2016guv,Khachatryan:2015yea,Khachatryan:2016sey,Sirunyan:2017guj,Sirunyan:2017tqo,Khachatryan:2015wka} have therefore place substantial effort in the search of Higgs pair production, in order to study the Higgs self coupling~\cite{Djouadi:1999rca,Dawson:1998py,Plehn:1996wb,Baur:2002rb,Baur:2002qd}. 
With an integrated high luminosity of $3000~{\rm fb}^{-1}$, it was shown the both $b\bar{b}\gamma\gamma$ and $b\bar{b}\tau\tau$ channels can be discovered at more than $5\sigma$ significance within the SM~\cite{Baglio:2012np}. 

The production of Higgs pair is sensitive to new physics, with obvious examples being non-standard Yukawa couplings
or Higgs self-coupling, the existence of new resonances, and  heavy coloured particles flowing in the loops~\cite{Contino:2010mh, Asakawa:2010xj, Dolan:2012ac, Chen:2014xra, Dawson:2015oha,Shen:2015pha,Etesami:2015caa,Cao:2015oaa,Chang:2016mso,Lim:2016ivn}.  
Higgs boson pair can also be produced in association with other particles in the decay of new resonances in new physics models. For example, a pair of the lightest supersymmetric particle (LSP) could generate a collider signal of Higgs boson pair with neutrinos in R-parity violating supersymmetry models in the case of Higgsino-like LSP~\cite{Biswas:2016ffy}. In certain models with extended gauge group and dark matter candidate, the heavy gauge boson $Z^\prime$ predominantly decay into a dark matter and a Higgs boson~\cite{Cheng:2003ju,Cheng:2004yc}.  If the $Z^\prime$ is produced only in pair due to a new parity  symmetry under which all SM particles are even  while $Z^\prime$ and dark matter are odd, the collider signature to search for $Z^\prime$ would be the Higgs boson pair plus large missing energy ($E_T^{miss}$) from dark matter. Furthermore, as the  $Z^\prime$ is much heavier than the Higgs boson and dark matter, the Higgs boson will be highly boosted, resulting in collimated decay products.  
In this study, we investigate in detail the associated  collider phenomenology at the LHC, taking Littlest Higgs Model with $T$-parity~\cite{Cheng:2003ju,Cheng:2004yc} as a benchmark model.  We employ jet substructure techniques \cite{Seymour:1993mx}
to exploit the boosted Higgs topology of the signal, resorting to the well-established  BDRS algorithm \cite{Butterworth:2008iy}.  

 The work is then organised as follows: In Sec.~\ref{sec:simp_model}, we give a brief introduction of the setup in the Littlest Higgs Model with $T$-parity. Subsequently, we consider the collider signatures of $pp\rightarrow (Z^\prime \rightarrow h \chi)(Z^\prime \rightarrow h \chi)$, where $\chi$ denotes dark matter at the 14 TeV High Luminosity LHC (HL-LHC) in Sec.~\ref{sec:collider}. The results are discussed in Sec.~\ref{sec:results}. The paper is then concluded by Sec.~\ref{sec:conclusions}.  

\section{Model} \label{sec:simp_model}

We begin by introducing the Little Higgs Model Model with $T$-parity as a benchmark model, in which heavy gague boson $Z'$ must be pair-produced. Subsequently, the $Z'$  decays exclusively into a Higgs boson and a dark matter candidate.
The prototype Littlest Higgs model (LH) \cite{ArkaniHamed:2002qy} interprets the Higgs boson as a pseudo-Goldstone mode in a nonlinear sigma model paramaterising the coset space $SU(5) /SO(5)$. The mass of Higgs boson is protected by the so-called collective symmetry breaking  mechanism up to one-loop order. As the global $SU(5)$ is broken down to $SO(5)$ at a certain scale $f$, the gauged symmetries $[SU(2)\times U(1)]_1\times [SU(2)\times U(1)]_2 \subset SU(5)$  are broken down to a diagonal $SU(2)\times U(1)\subset SO(5)$, which is identified as the SM $SU(2)_L\times U(1)_Y$.

$T$-parity as a $\mathbb{Z}_2$ symmetry\footnote{In a way $T$-parity resembles the $R$-parity in SUSY, except that cancellation of Higgs mass correction is achieved by particles of the opposite spin-statistics in SUSY, but that of same spin-statistics here.} between the two $SU(2)\times U(1)$ copies  is imposed by demanding $g_1=g_2 = \sqrt{2}g$ and $g'_1=g'_2 = \sqrt{2}g'$, where $g_{1,2} (g^\prime_{1,2})$ and $g (g^\prime)$ are the gauge couplings of $SU(2)_{1,2}(U(1)_{1,2})$ and  the SM $SU(2)_L(U(1)_Y)$, respectively. The SM gauge fields, $W_\mu^a$  may be recovered from linear combinations of $SU(2)_{1,2}$ gauge fields $W^a_{\mu 1} $ and $ W^a_{\mu 2}$, with $a\in\{1,2,3\}$, whilst combinations that that are orthogonal to the SM configurations are realised as the heavy gauge bosons $W^{a\prime}_{\mu}$. Similar setup happens in $U(1)$ as well. As a result, one can easily find that heavy gauge bosons carry charge $-1$ ($T$-odd), and SM fields $+1$ ($T$-even) under $T$-parity transformation. 
The immediate implication of such parity is that the constraints from electroweak observations can be significantly alleviated by forbidding mixing between heavy and the SM gauge bosons.  Furthermore, the $T$-odd partner of photon is suitable for dark matter candidate \cite{Asano:2006nr,Birkedal:2006fz,Wang:2013yba,Chen:2014wua} whose mass is now bounded below by $m_h/2$ \cite{Wu:2016rwz,Reuter:2012sd,Yang:2014mba,Han:2014qia,Han:2013ic}.

After the electroweak symmetry is broken, the masses of $T$-odd heavy gauge bosons are given by :  
\begin{equation}
  m_{Z^{\prime}}\simeq m_{W^{\prime\pm}}= gf\left( 1-\frac{v^2}{8f^2} \right), \quad    m_{A^\prime}= \frac{g\prime f}{\sqrt{5}}\left( 1-\frac{5v^2}{8f^2} \right),
\end{equation}
where $A^\prime$ is the dark matter candidate. 
Due to $T$-parity exchanging $SU(2)_1$ and $SU(2)_2$, it is necessary to introduce two fermion doublets $\psi_{1,2}$ that transform linearly under the respective groups. In the end, we have heavy fermions with masses given by
\begin{equation}
  m_{u'}=\sqrt{2}\kappa_q f\left( 1-\frac{v^2 }{8f^2} \right),~  m_{d'}=\sqrt{2}\kappa_q f,~
 m_{\nu'}=\sqrt{2}\kappa_\ell f\left( 1-\frac{v^2 }{8f^2} \right),~  m_{\ell'}=\sqrt{2}\kappa_\ell f,
\end{equation}
where $u'$, $d'$, $\nu'$ and $\ell'$ are $T$-odd partners of SM up-type quarks, down-type quarks, neutrinos and charged leptons, respectively. The $\kappa_q$ and $\kappa_\ell$ denote the Yukawa-type coupling strength. The couplings between $Z'$ and SM quarks are generated in the gauge interactions as follows:
\begin{equation}
 {\cal L} \supset \frac{g}{2} \left( 1+\mathcal{O} \left(\frac{v^2}{f^2}\right)\right)Z'^\mu \bar{u}_L\gamma_\mu u'_L-\frac{g}{2} \left( 1+\mathcal{O} \left(\frac{v^2}{f^2}\right)\right)Z'^\mu \bar{d}_L\gamma_\mu d'_L+h.c..
  \label{eq:model}
\end{equation}
In the case that $T$-odd fermions are heavier than $Z'$, the only channel allowed for $Z'$ decay is $Z'\to h A'$. 
The LHT provides a framework whereby a heavy gauge $Z'$ conveniently couples the Higgs and dark sectors. The collider phenomenology of the $Z'$  is studied in the subsequent parts of this paper without assuming its realisation via any specific model. This interpretation is made possible by treating the mass of dark matter $m_{A'}$ and coupling of $Z'$ to SM quarks and heavy quarks in Eq.~(\ref{eq:model}) as free parameters. Also, we assume that the heavy quarks ($u'$ and $d'$) are much heavier than $Z'$.

\section{Collider study} \label{sec:collider}

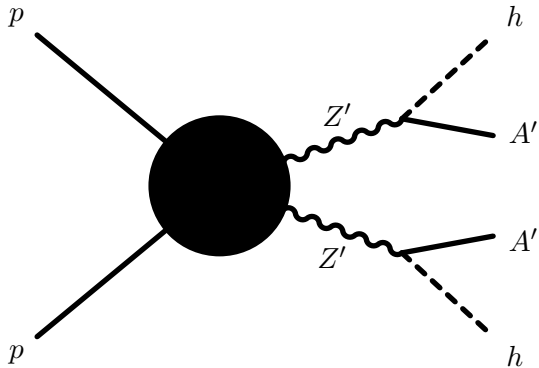
\begin{figure}[h!]
\vspace{4ex}
\center
\unitlength=2mm
\begin{fmffile}{feyn}						
\begin{fmfgraph*}(30,20)
\fmfstraight
\fmfpen{2}
\fmfleft{p1,p2}   \fmflabel{$p$}{p1}    \fmflabel{$p$}{p2}  
\fmfright{h1,a1,a2,h2}  \fmflabel{$h$}{h1}    \fmflabel{$h$}{h2} 
\fmflabel{$A'$}{a1}    \fmflabel{$A'$}{a2}
\fmf{vanilla}{p1,v1,p2}
\fmf{boson,label=$Z'$ }{v1,z1}
\fmf{boson,label=$Z'$}{v1,z2}
\fmf{dashes}{z1,h1}
\fmf{dashes}{z2,h2}
\fmf{vanilla}{z1,a1}
\fmf{vanilla}{z2,a2}
\fmfv{decor.shape=circle,decor.size=0.3w}{v1}
\end{fmfgraph*}
\end{fmffile}
\caption{Double Higgs production associated with large $E_T^{miss}$ in the LHT model. The missing transverse energy originates from $A'$ as in (\ref{eq:prod_mode}) } 
  \label{fig:feyn}
\end{figure}

The LHT model provides a means of $2h+E_T^{miss}$ production through the decay of $Z'$. The missing transverse energy  component is attributed to the dark matter. The relevant channel of production is    
\begin{equation} \label{eq:prod_mode}
  pp\rightarrow{Z'} {Z'}\rightarrow h (\rightarrow \gamma\gamma) A' h(\rightarrow b\overline{b} ) A' ,
\end{equation}
where $A'$ denotes the dark matter, as seen in Fig~\ref{fig:feyn}. Here we focus on the $b\bar{b}\gamma\gamma$ final state as two Higgs bosons decay at  the 14 TeV HL-LHC. 
%
The dominant backgrounds to the signal process in order of contribution are as follows:
\begin{itemize}
  \item[(B1)] $t(\rightarrow \ell ^+{\nu} b)\bar{t}(\rightarrow \ell^- \overline{\nu} \overline{b}) \gamma$ --- this reducible background may mimic the signal if one of lepton is missed and the other is misidentified as a photon.
%
  \item[(B2)] $t(\rightarrow \ell ^+{\nu} b)t(\rightarrow \ell^- \overline{\nu} \overline{b}) \gamma\gamma$  --- this becomes a background if both leptons are missed ,or if one photon is missed and the lepton is misidentified as a photon. Both this and (B1) contain large $E_T^{miss}$ signatures. However, the production of photon and $b\overline{b}$ pairs are  non-resonant and therefore expected to be effectively suppressed via a succession of mass window cuts. 
  \item[(B3)] $t\overline{t} h(\rightarrow \gamma\gamma$) --- same as (B2), except the diphoton is produced via a resonant Higgs. Although this is suppressed by the branching ratio, it becomes comparable with (B2) since the diphoton invariant mass cut for  is unable to distinguish signal against background.    
  \item[(B4)] $Z(\rightarrow b\overline{b})h(\rightarrow \gamma\gamma$) ---
   This background will be effectively suppressed by an $E^{miss}_T$ cut. Since the mass of $Z$ is sufficiently close to that of the Higgs, the effect of mass window cuts will not be very effective.     
\end{itemize}
 The backgrounds  $ jj\gamma\gamma$, $b\bar{b}\gamma\gamma$, $jjh$ and $b\bar{b}h$ have been studied and were found to be negligible after $E^{miss}_T$ cut.
%


The effective Lagrangian in Eq. (\ref{eq:model}) was implemented by \texttt{FeynRules\_v2.3} \cite{Alloul:2013bka} with SM parameters taken from \cite{Nakamura:2010zzi}. The signal and background matrix elements were generated by \texttt{MadGraph5\_aMC@NLO} package \cite{Alwall:2014hca}  with default parton level cuts, and convolved with the CTEQ6L parton distribution function \cite{Pumplin:2002vw} using default dynamical renormalisation ($\mu_R$) and factorisation ($\mu_F$) scales. Parton showering was subsequently performed by  \texttt{Pythia} \cite{Sjostrand:2007gs} followed by 
\texttt{Delphes-3} \cite{deFavereau:2013fsa} for detector simulation, where the (mis-)tagging efficiencies and fake rates assume their default values, with the exception of the angular resolution of photons which is set to $0.3$. 

\section{Results} \label{sec:results} 
\begin{figure}[h!]
  \centering
  \includegraphics[width=.8\textwidth,clip=true,trim=0mm 0mm 0mm 0mm]{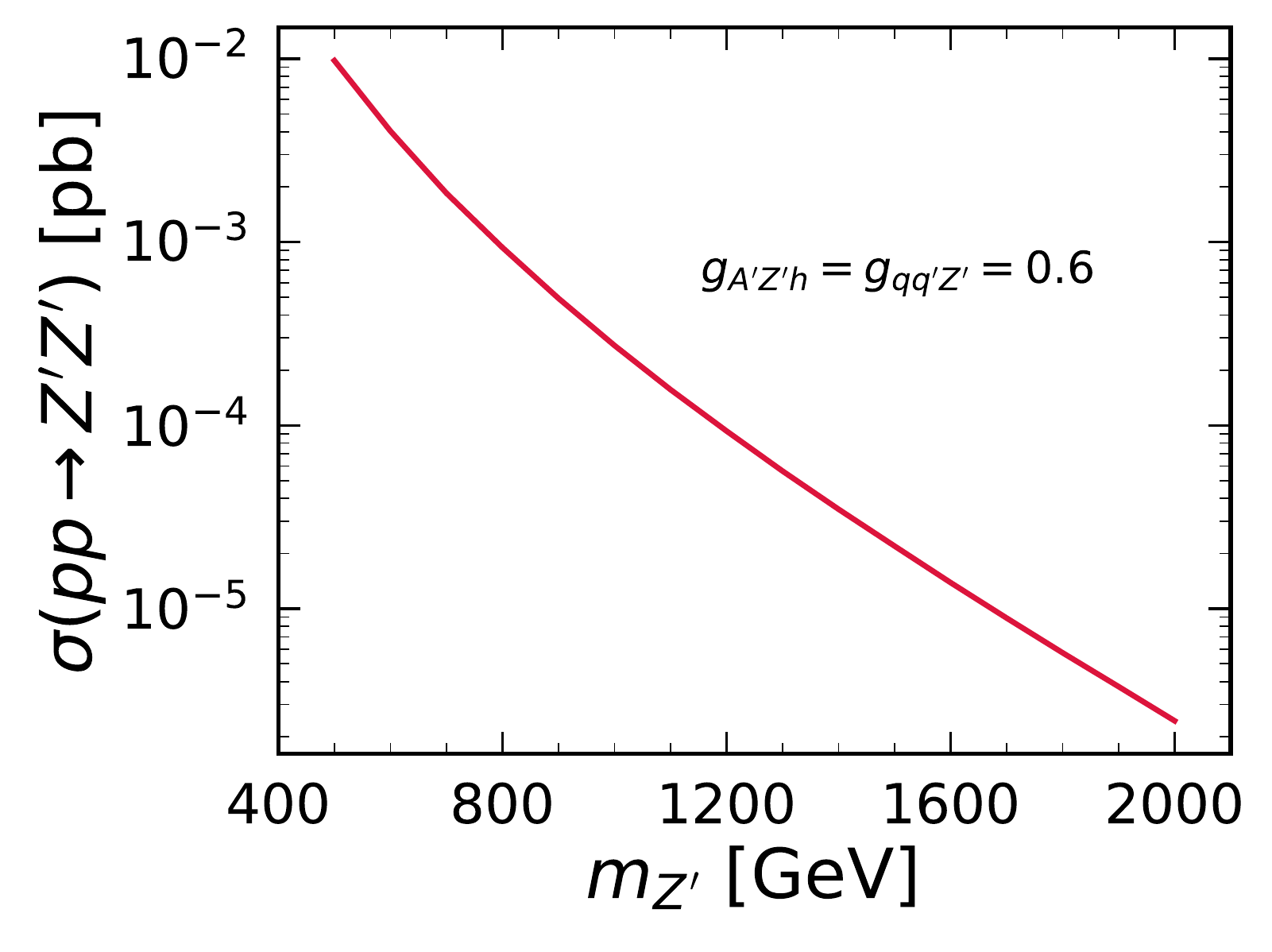}
  \caption{Cross section as a function of the $m_{Z'}$ mass at the 14 TeV LHC.  A 100\% decay rate has been assumed for $Z' \rightarrow A' h $.} 
  \label{fig:crossX}
\end{figure}

Fig.~\ref{fig:crossX} shows the cross section where $g=0.6$ is adopted in (\ref{eq:model}), as well as the masses of heavy quarks $m_{u'}=m_{d'}=2.5$ TeV.
The increase mass of $Z'$ makes it more difficult to be pair produced, as reflected by the decreasing cross section. As mentioned previously, the $Z'$ predominantly decays via $Z'\rightarrow h A'$ with a branching ratio $100\%$.

The signal analysis (cf. Table~\ref{tab:14Tev_cutflow}) begins with the basic selection criteria (C0) on transverse momenta and rapidities:
\begin{equation}
  \begin{aligned}
    p_T^\gamma > 15~\text{GeV}, \quad | \eta^\gamma | < 2.5\\ 
  \end{aligned}
\end{equation}
which reflects the triggering capabilities and detector coverage at the LHC. 
We demand two or more photons ($N_\gamma \geq 2$) to reconstruct one of the Higgs bosons. For the  Higgs boson decaying into two $b$-quarks, we focus on the boosted regime and adopt jet substructure technique. 

\begin{figure}[h!]
  \centering
  \includegraphics[width=.8\textwidth,clip=true,trim=0mm 0mm 0mm 0mm]{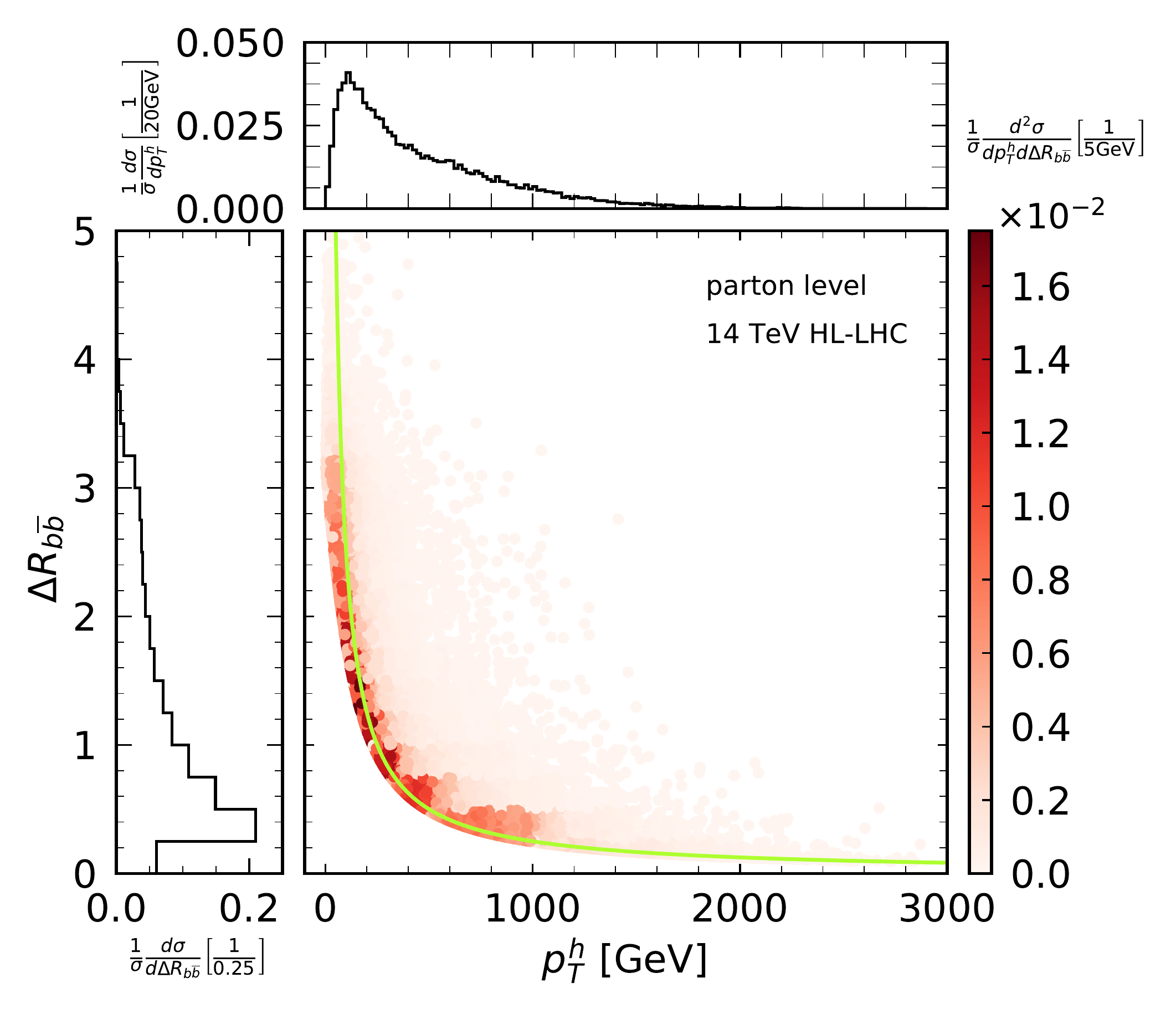}%
  \caption{The correlation of transverse momentum of the Higgs boson and $\Delta R$ separation between two bottom quarks in its decay. The green line shows the relation (\ref{eq:boost}).
 }
  \label{fig:dR_pt_prtlvl_}
\end{figure}

Fig.~\ref{fig:dR_pt_prtlvl_} shows a parton level simulation of the $\Delta R$ of the 2$b$ decay of the Higgs the signal events. The  separation finds good agreement with:  
\begin{equation}
  \Delta R(b,\overline{b}) \sim \frac{2m_h }{p_T^h} \label{eq:boost}
\end{equation}
A boost regime of $p_T> 200$ GeV corresponds to $\Delta R\lesssim 1.25$. This motivates the setting of the cone radius to $R=1.2$  in order to  cluster  fatjets  via the Cambridge/Aachen algorithm  \cite{Dokshitzer:1997in,Wobisch:1998wt}.
Jet substructure was subsequently used to tag Higgs-candidate fatjet by utilising the BDRS  algorithm \cite{Butterworth:2008iy} made available by \texttt{fastjet} \cite{Cacciari:2011ma} .  The BDRS algorithm iteratively reverses the clustering for a jet $j$ from constituents $j_1$ and $j_2$, until a significant mass drop is observed and without the splitting being too large. The parameters for mass drop $\mu$ and splitting $y$ are chosen to be:
\begin{equation}
  \mu =\frac{\max(m_{j_1},m_{j_2}) }{m_j}<  0.667
\end{equation}
\begin{equation}
  y=\frac{\min(p^2_{T,j_1},p^2_{T,j_2})}{m^2_j}\Delta R^2(j_1,j_2)>  0.09 
\end{equation}
Subsequently, the subjets are reclustered with $R_\text{filt} = \min\left( 0.3 , R_{b\overline{b}}/2 \right)$, keeping only three hardest subjets, imposing a minimum of $p_T^{min} = 50$ GeV on the transverse momentum. 
$b$-tagging is achieved with a probability 70\% if a parton level $b$ quark ($p_T^b>20$ GeV $|\eta_b|<2.5$) is found within the cone radius of $\Delta R = 0.3$ around jet direction. In order to retain as much of the signal as possible, we require only one $b$-tag.

\begin{figure}[h!]
  \centering
  \includegraphics[width=.8\textwidth,clip=true,trim=0mm 0mm 0mm 0mm]{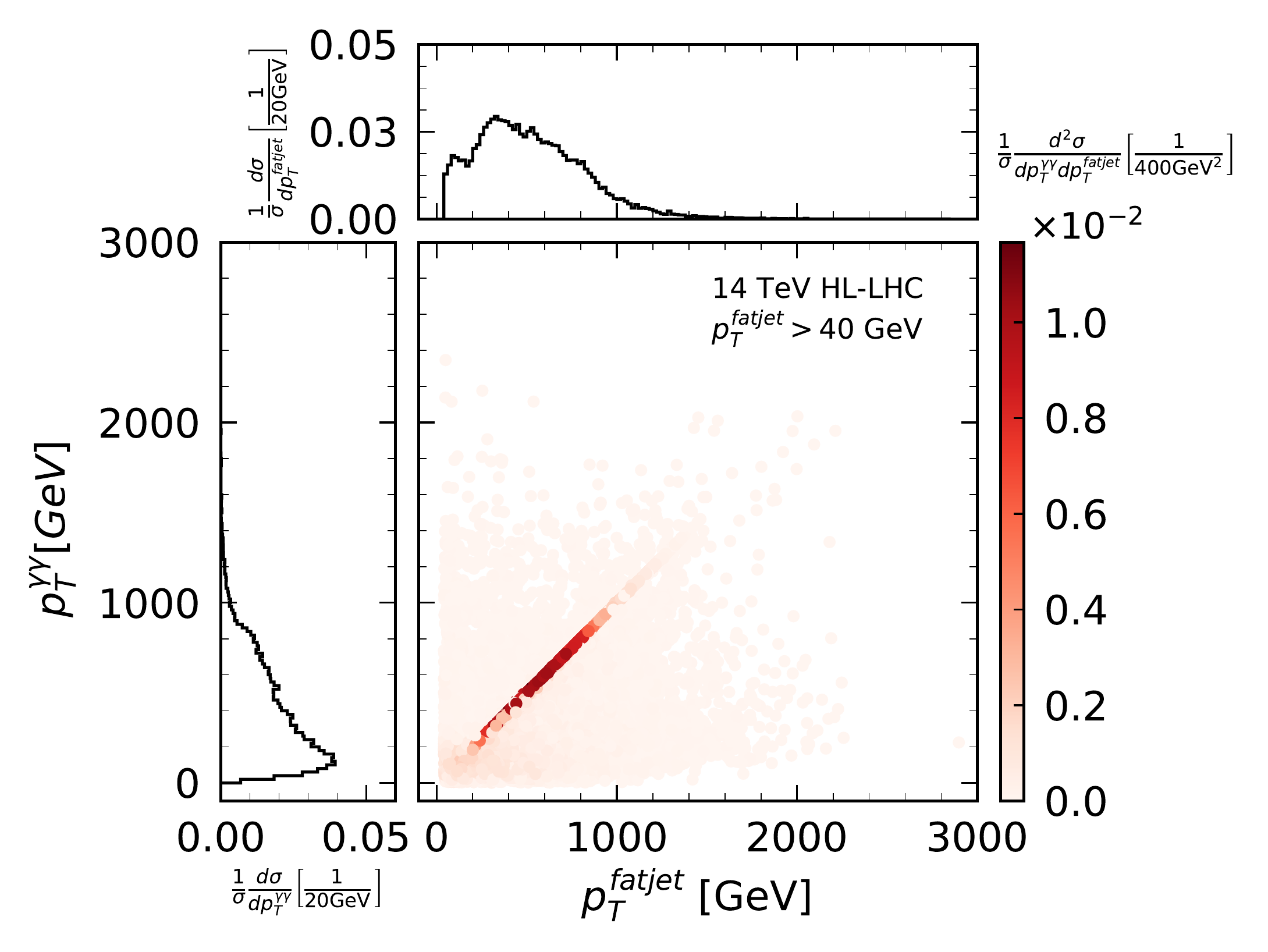}%
  \caption{The transverse momentum correlation between two Higgs candidates as reconstructed by $2\gamma$ and the BDRS algorithm without $b$-tagging.}
  \label{fig:reclvl_pt_pt}
\end{figure}
\begin{figure}[h!]
  \centering
  \includegraphics[width=.7\textwidth,clip=true,trim=0mm 5mm 0mm 0mm]{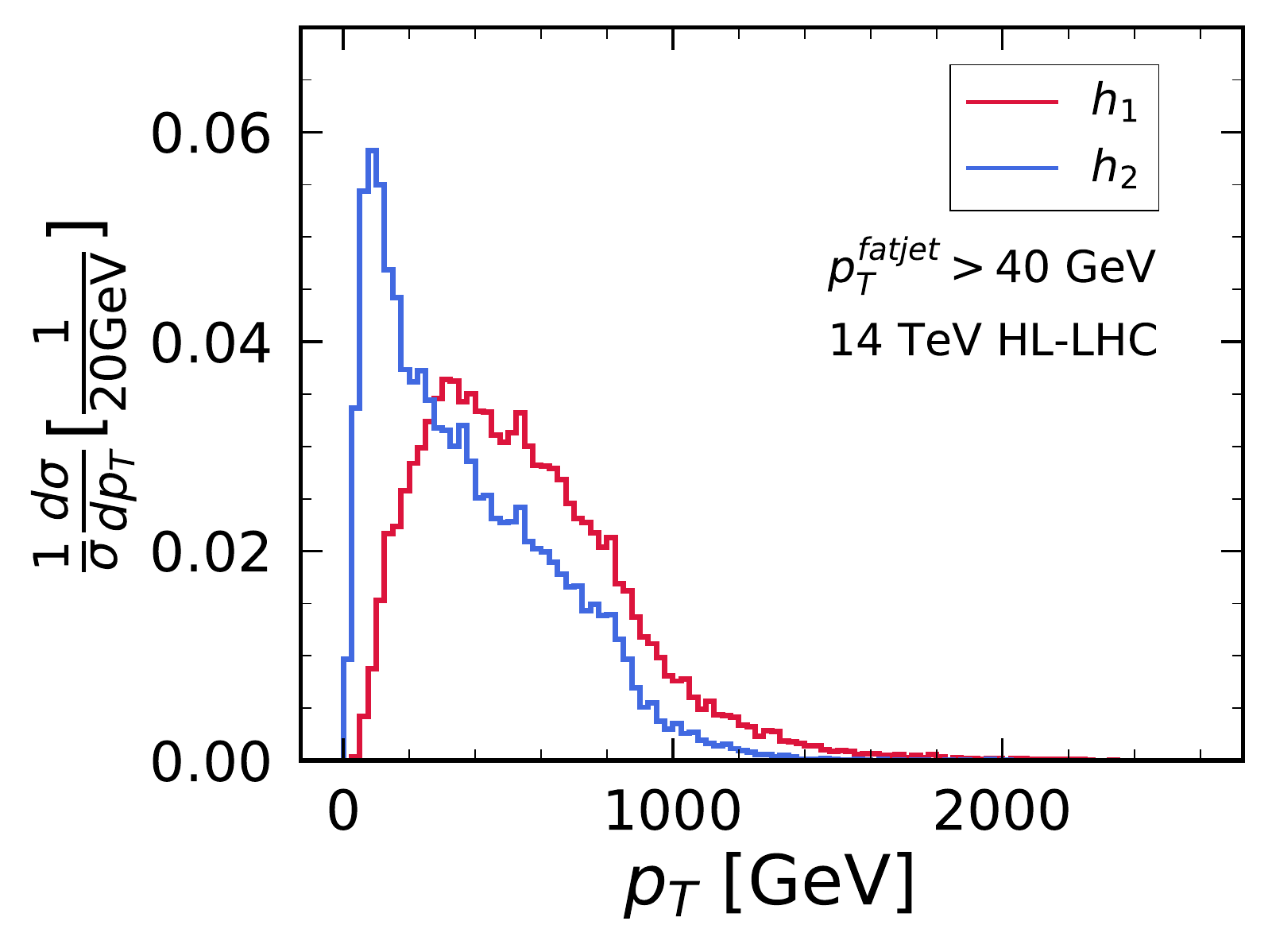}%
  \caption{The transverse mometum of the Higgs candidate where $p_T^{h_{1}} :=\max(p_T^{fatjet},p_T^{\gamma\gamma})$ and  $p_T^{h_{2}} :=\min(p_T^{fatjet},p_T^{\gamma\gamma})$. }
  \label{fig:reclvl_pt1_pt1}
\end{figure}
\begin{figure}[t!]
  \centering
   \includegraphics[width=.49\textwidth,clip=true,trim=0mm 0mm 0mm 0mm]{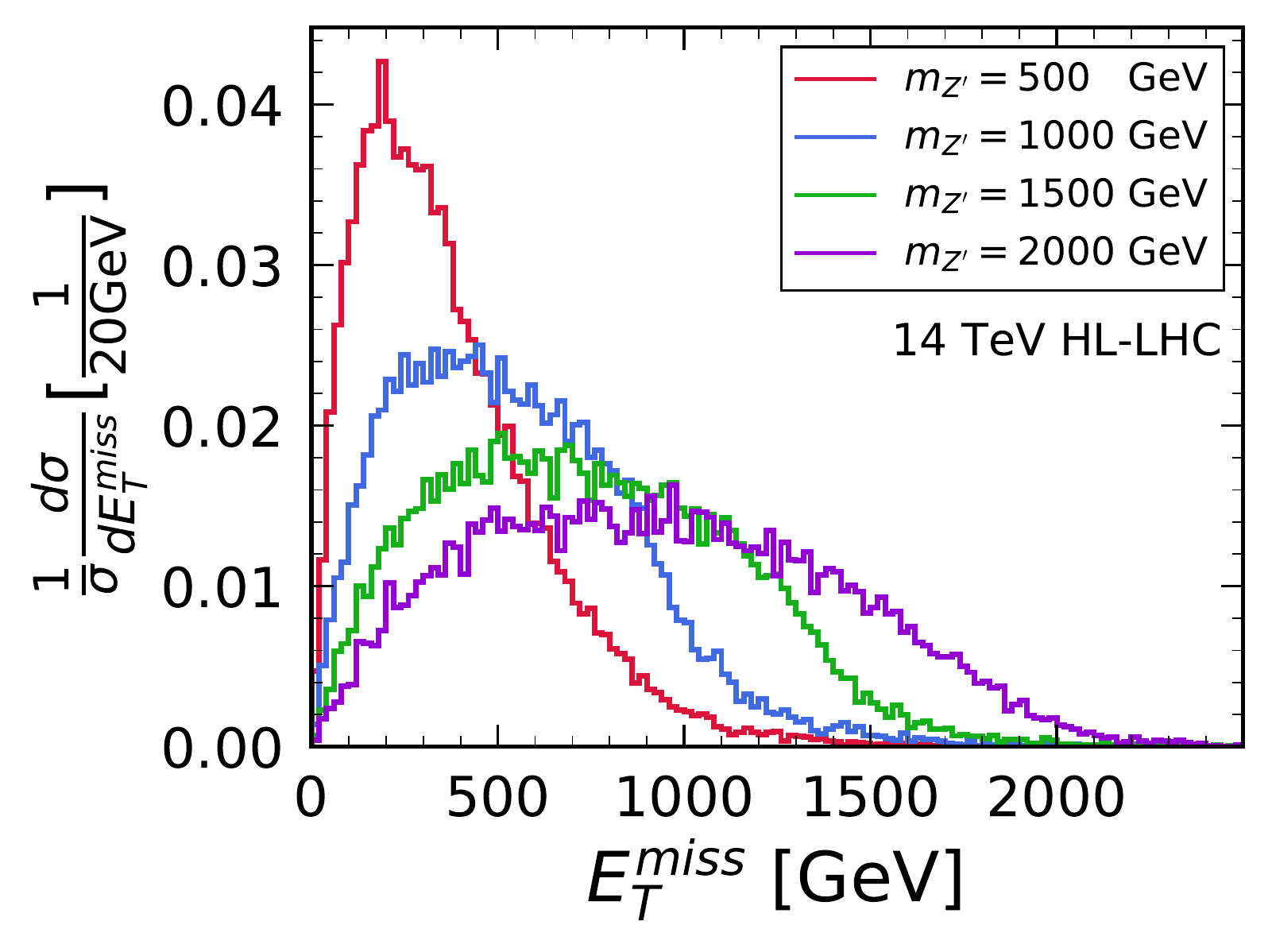}%
    \includegraphics[width=.49\textwidth,clip=true,trim=0mm 0mm 0mm 0mm]{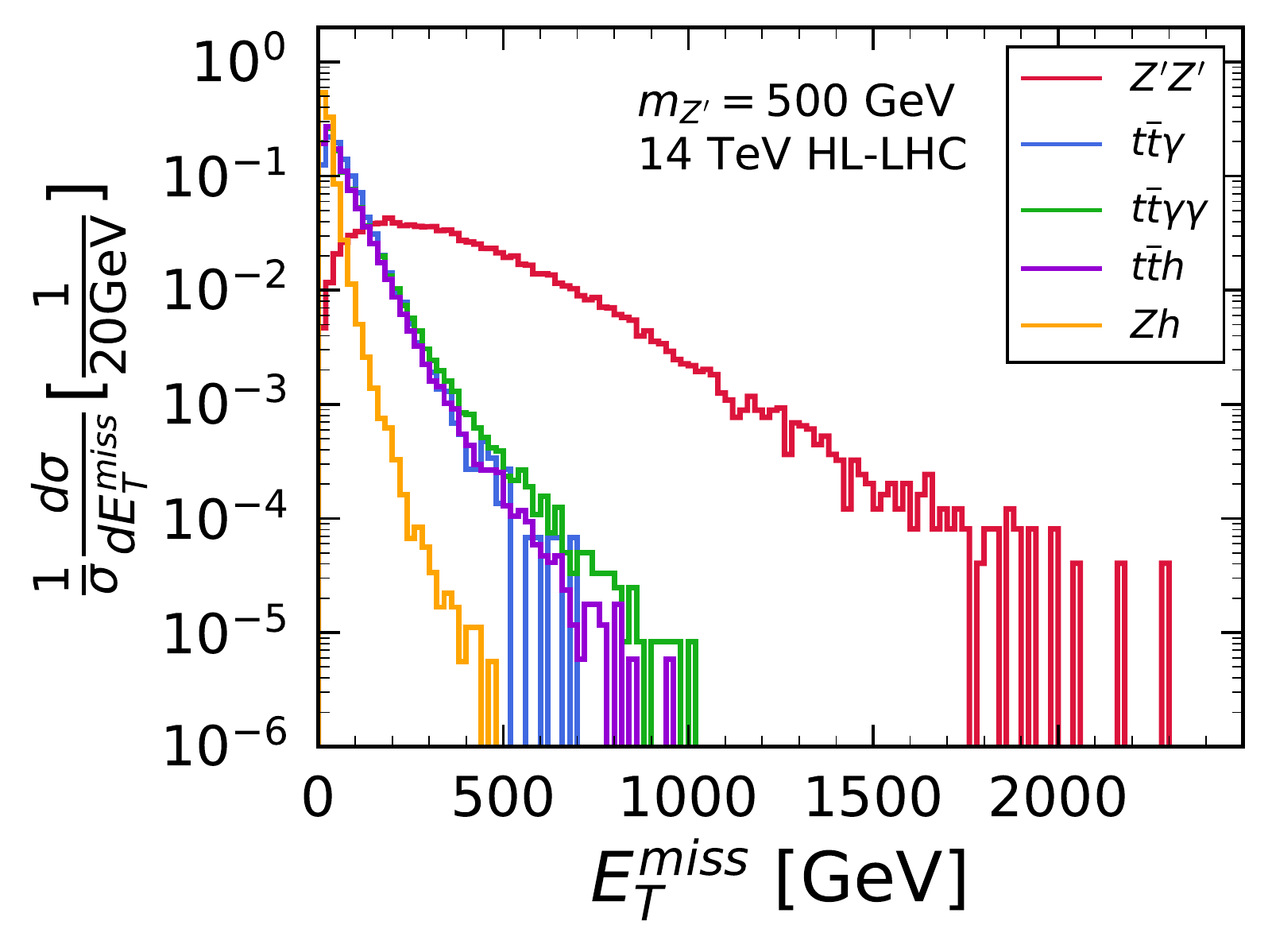}%
    \caption{The $E_T^{miss}$ for signals assuming different masses of $m_{Z'}$ (left)  and  signal versus various backgrounds.} 
  \label{fig:MET}
\end{figure}

Fig.~\ref{fig:reclvl_pt_pt} shows the transverse momentum of the Higgs boson candidates reconstructed via the algorithm above. It suggests symmetric energy distribution between the two  $h$ bosons that are daughters in subsequent decays of heavy $Z'$ bosons. After $p_T$ ordering, one can see from Fig.~\ref{fig:reclvl_pt1_pt1} that the leading Higgs boson becomes sufficiently boosted. 
From the left panel of Fig.~\ref{fig:MET}, we show the distributions of  $E_T^{miss}$ for signal as well as the main backgrounds. It is evident that $E_T^{miss}$, mainly due to the undetected dark matters $A'$, gets more energetic  with increasing $m_{Z'}$. Even though $t\overline{t}+X$ backgrounds  contribute to the $E_T^{miss}$ spectrum, the right panel of Fig.~\ref{fig:MET} shows that it tends to be significantly softer than that of the signal. 
The neutrinos which account for the $E_T^{miss}$ in the background events are produced via charged weak current, and hence are accompanied by charged leptons. These leptons are identified in the range $p^\ell_T >20$ GeV, $|\eta^\ell|<2.5$. 
 We see that they can be effectively suppressed at least an order of magnitude using lepton veto (C3) in conjunction with a $E_T^{miss}>150$ GeV cut (C4).  
For $jjh$, the missing transverse energy arise completely due to detector inefficiencies and so is removed around three orders of magnitude due to (C4).  

\begin{figure}[h!]
  \centering
  \includegraphics[width=.5\textwidth,clip=true,trim=0mm 0mm 0mm 0mm]{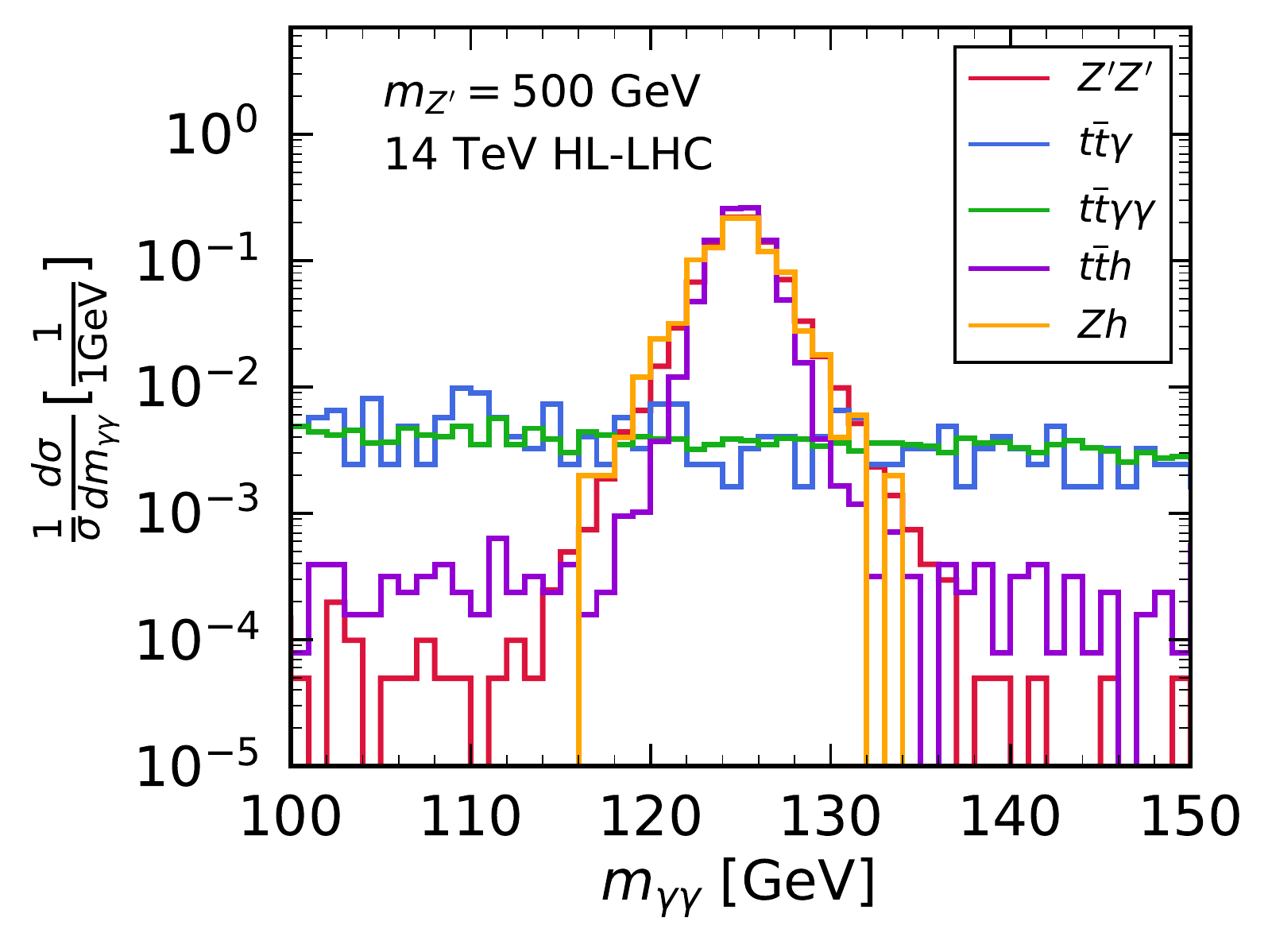}%
  \includegraphics[width=.5\textwidth,clip=true,trim=0mm 0mm 0mm 0mm]{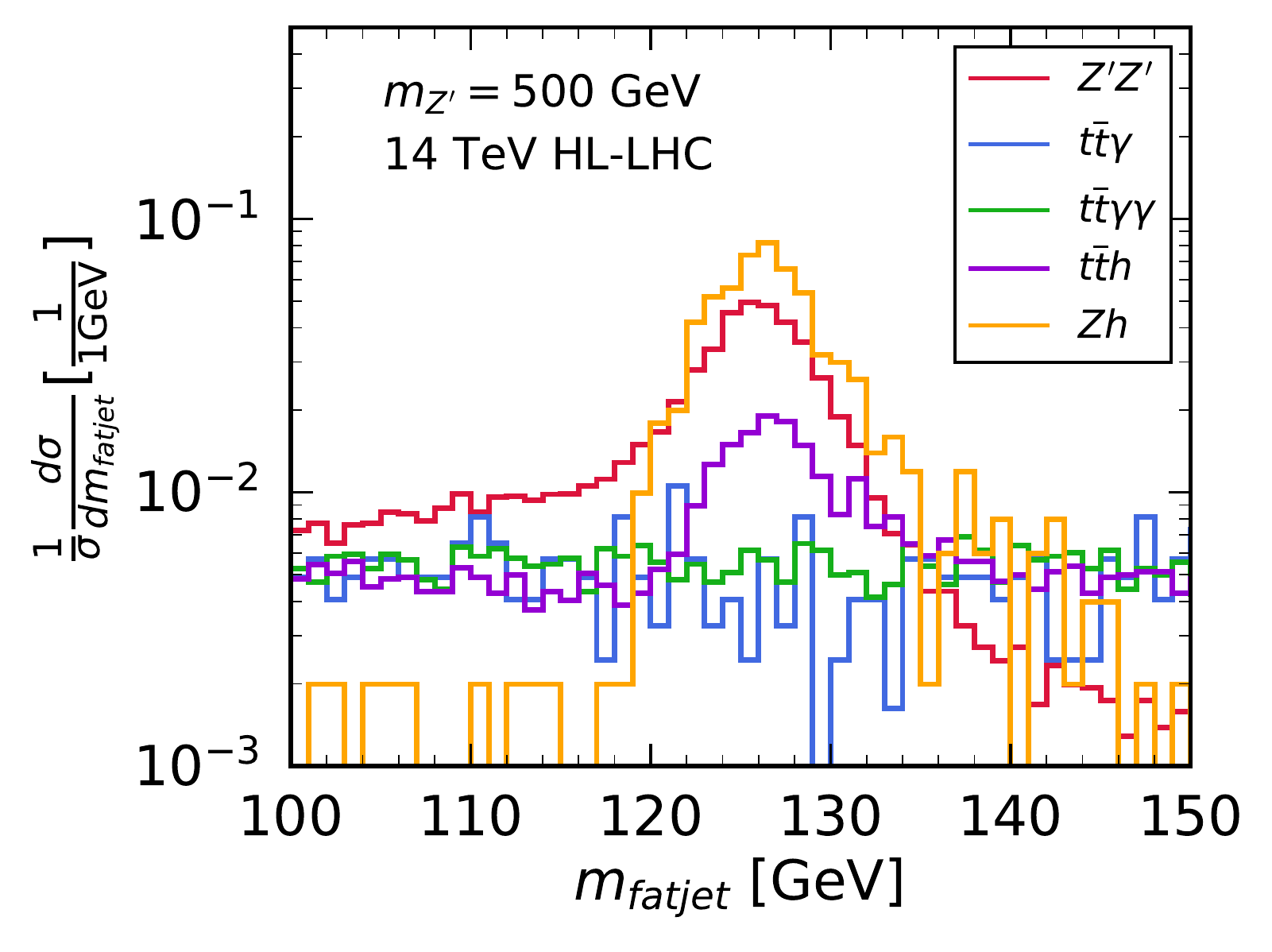}%
  \caption{The reconstructed Higgs invariant mass $m_h$ from   $\gamma\gamma$ (left)  and fatjet (right) before cuts (C4). 
  } 
  \label{fig:m_h}
\end{figure}

In order to take care of the continuum process (B1) and (B2), we employ mass window cuts (C5) and (C6) around $m_h=125$ GeV  on the diphoton and fatjet mass respectively. This results in two orders of magnitude reduction for these backgrounds, but simultaneously removes $\sim 1/3$ of the signal events. On the other hand, for the $jjh$ and $t\overline{t}h$ backgrounds,  two photons are mainly generated in Higgs decay and therefore show a peak structure in $m_{\gamma\gamma}$ in Fig.~\ref{fig:m_h}. This renders the diphoton invariant mass cut (C5) less effective. Instead, one relies on the fatjet invariant mass cut (C6) to reduce the background by a factor of $\sim3$, whilst that of the signal is reduced by $\sim 1/3$. For the $Zh$ background, although the $h\rightarrow \gamma\gamma$ decay is irreducible, and the invariant mass of the fatjet due to $Z\rightarrow b\overline{b}$ is sufficiently close to that of the Higgs boson, the initial cross section of the process is low enough such that the $E_T^{miss}$ selection cut suffices to bring this background in control.

Given the cut efficiencies, we can extend the results summarised in Tab.~\ref{tab:14Tev_cutflow} to project the sensitivity of 14 TeV HL-HLC with $m_{Z'}=500~{\rm GeV}$ and $m_{A'}=100~{\rm GeV}$, assuming fixed couplings $g_{qq'Z'}=g_{hZ' A'}=0.6$.  We also show the comparison between there benchmark scenarios with $m_{Z'} \in \{500,1000, 2000\}$ GeV in Appendix. The LHC sensitivity is summarised in Fig.~\ref{fig:Sig} by the approximate median significance (AMS)\cite{Cowan:2010js}:%
\begin{equation}
  \mathcal{S} := \sqrt{2\left( (s+b) \ln \left( 1+\frac{s}{b} \right)-s \right)}=\frac{s}{\sqrt{b}}\left( 1+\mathcal{O}\left(\frac{s}{\sqrt{b}}\right)  \right)
\end{equation}
which characterises the observed statistical significance for rejecting the  SM-only hypothesis.  Assuming a coupling strength of 0.6, the discovery is expected with the luminosity of 100 fb$^{-1}$ for a $Z'$ of $500$ GeV in mass, corresponding to $\sim11$ events at 14 TeV LHC. For different couplings of $g_{qq'Z'}$, the needed luminosity for discovery  of a $500$ GeV $Z'$ is also shown in the right panel  of Fig.~\ref{fig:Sig}.

\begin{figure}[h!]
  \centering
  \includegraphics[width=.5\textwidth,clip=true,trim=0mm 0mm 0mm 0mm]{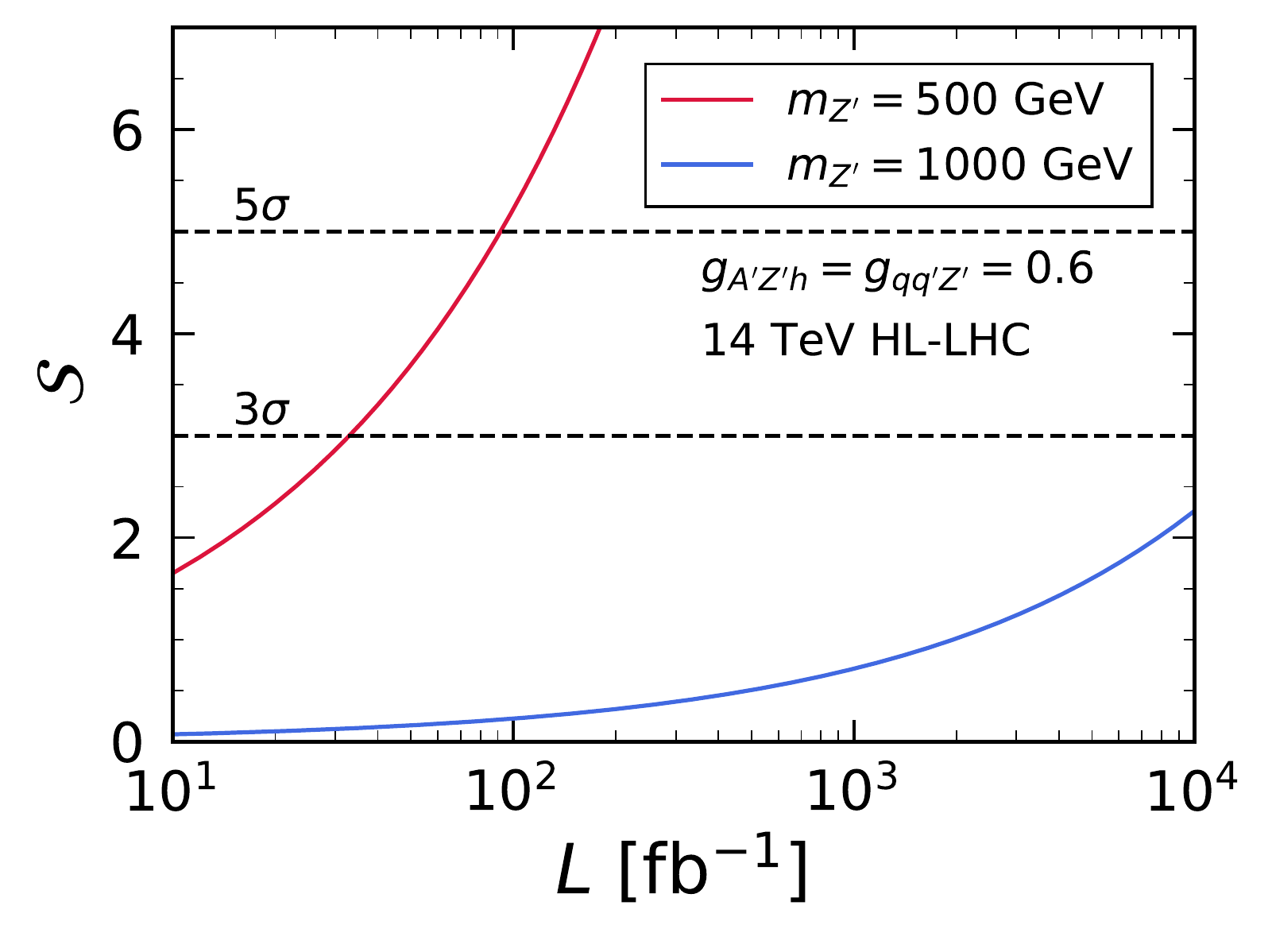}%
  \includegraphics[width=.5\textwidth,clip=true,trim=0mm 0mm 0mm 0mm]{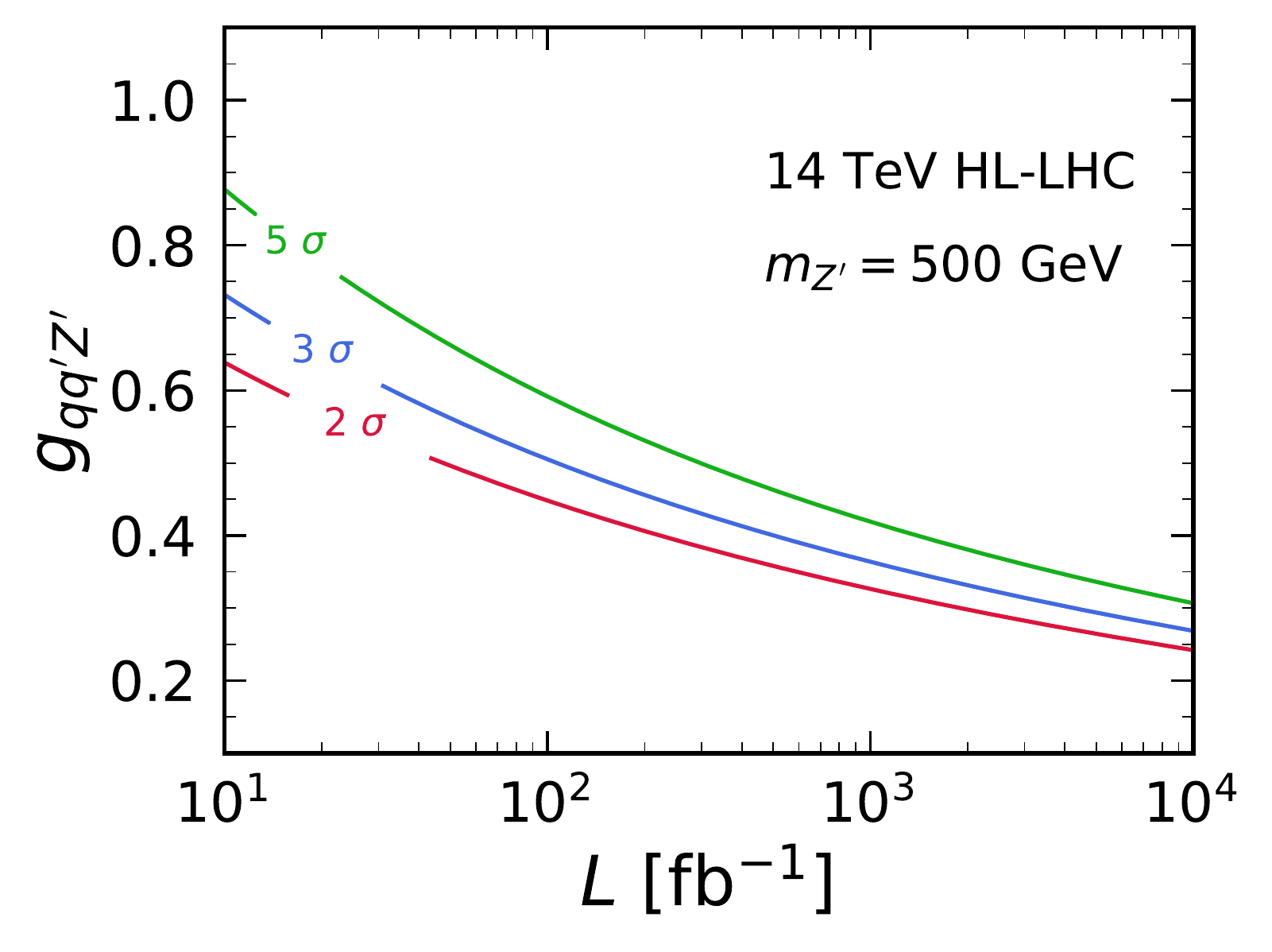}
  \caption{The discovery potential of $pp \to Z'Z'\to h h A'A'$ at the 14 TeV LHC.\label{fig:Sig}}
  
\end{figure}

\begin{landscape}
\begin{centering}
\begin{table}[p]
\fontsize{10pt}{12pt}\selectfont
\begin{center}

{\renewcommand{\arraystretch}{1.25}\newcolumntype{C}[1]{>{\centering\let\newline\\\arraybackslash\hspace{0pt}}m{#1}} 
\begin{tabular}{ccc|cc|c|c|c|c|c}
\hline 
\multicolumn{3}{c|}{\multirow{3}{*}{Cuts}} & \multicolumn{7}{c}{ $\sigma$ [$10^{-4}$ fb] } \\
\cline{4-10} 
& & & \multicolumn{2}{c|}{ $pp\rightarrow Z' Z' $}
& 
{ $pp\rightarrow t\bar{t}\gamma$ } &
{ $pp\rightarrow t\bar{t}\gamma\gamma $} &
{ $pp\rightarrow jjh(\rightarrow \gamma\gamma)$ } &
{ $pp\rightarrow t\bar{t}h(\rightarrow\gamma\gamma)$ } &
{ $pp\rightarrow Z(\rightarrow b \overline{b}) h(\rightarrow\gamma\gamma)$ } \\

\hline
(C0) & \multicolumn{2}{c|}{Basic cut}  & \multicolumn{2}{c|}{$9.70\times 10^3$}
& 2.44 $\times 10^{7}$ & 1.18 $\times 10^{5}$ & 5.74 $\times 10^{4}$ & 7.55 $\times 10^{3}$ & 1.41 $\times 10^{3}$ \\
\hline
(C1) & \multicolumn{2}{c|}{ $N_\gamma \ge 2$ } & \multicolumn{2}{c|}{$6.10\times 10^3$ }
& 5.40 $\times 10^{5}$ & 4.61 $\times 10^{4}$ & 3.23 $\times 10^{4}$ & 4.18 $\times 10^{3}$ & 6.80 $\times 10^{2}$ \\
\hline
(C2) & \multicolumn{2}{c|}{ $N_\text{fatjet} \ge 1$ } & \multicolumn{2}{c|}{$6.01\times 10^3$}
& 5.37 $\times 10^{5}$ & 4.60 $\times 10^{4}$ & 2.95 $\times 10^{4}$ & 4.17 $\times 10^{3}$ & 6.33 $\times 10^{2}$ \\
\hline
(C3) & \multicolumn{2}{c|}{ $N_\ell\leq  0$ } & \multicolumn{2}{c|}{$6.00\times 10^3$}
& 4.46 $\times 10^{5}$ & 3.54 $\times 10^{4}$ & 2.95 $\times 10^{4}$ & 3.22 $\times 10^{3}$ & 6.32 $\times 10^{2}$ \\
\hline
(C4) & \multicolumn{2}{c|}{ $E_T^{miss} \geq 150~\text{GeV}$ } & \multicolumn{2}{c|}{$4.91\times 10^3$}
& 3.89 $\times 10^{4}$ & 3.12 $\times 10^{3}$ & 4.12 $\times 10^{1}$ & 2.40 $\times 10^{2}$ & 1.77 \\
\hline
(C5) & \multicolumn{2}{c|}{  $|m_{\gamma\gamma} -125~\text{GeV}|< 10 \ \text{GeV}$ } & \multicolumn{2}{c|}{$4.84\times 10^3$}
& 2.81 $\times 10^{3}$ & 2.29 $\times 10^{2}$ & 4.06 $\times 10^{1}$ & 2.27 $\times 10^{2}$ & 1.77 \\
\hline
(C6) & \multicolumn{2}{c|}{ $|m_{\text{fatjet}}-125~\text{GeV}|< 30 \ \text{GeV}$ } & \multicolumn{2}{c|}{$3.23\times 10^3$}
& 8.55 $\times 10^{2}$ & 7.97 $\times 10^{1}$ & 2.51 $\times 10^{1}$ & 8.90 $\times 10^{1}$ & 1.23 \\
\hline
(C7) & \multicolumn{2}{c|}{ (1+)$b$ in fatjet } & \multicolumn{2}{c|}{$1.24\times 10^3$    }
& 1.83 $\times 10^{2}$ & 3.47 $\times 10^{1}$ & - & 2.42 $\times 10^{1}$ & 4.23 $\times 10^{-2}$ \\
\hline
(C8) & \multicolumn{2}{c|}{ $m_{hh} > 252\ \mathrm{GeV}$ } & \multicolumn{2}{c|}{$1.22\times 10^3$}
& 1.83 $\times 10^{2}$ & 3.41 $\times 10^{1}$ & - & 2.31 $\times 10^{1}$ & 3.53 $\times 10^{-2}$\\
\hline
(C9) & \multicolumn{2}{c|}{ $p_T^\text{fatjet} >200~\mathrm{GeV}$ } & \multicolumn{2}{c|}{$1.20\times 10^3$}
& 1.83 $\times 10^{2}$ & 3.27 $\times 10^{1}$ & - & 2.07 $\times 10^{1}$ & 3.17 $\times 10^{-2}$\\
\hline\hline
\end{tabular}}

\caption{ Select cuts and resulting cross sections for the signals of different masses  14 TeV LHC. Here, we set $m_{Z'}=500~{\rm GeV}$, $m_{A'}=100~{\rm GeV}$, and $g_{Z' A' h} = g_{uu'Z'}=0.6$. \label{tab:14Tev_cutflow}}

\end{center}
\end{table}
\end{centering}
\end{landscape}

\section{Conclusions} \label{sec:conclusions} 
It is conceivable that the dark sector may indeed play a role in stablising the Higgs boson mass. Yet, dark matter candidates are usually stablised via a discrete symmetry of some sort. Should such dark matter  be stablisied by being odd under a $\mathbb{Z}_2$ symmetry and couples to the Higgs boson via a  heavy resonance, it is clear that enforcing that SM particles are even forces this resonance to carry odd-parity under the same symmetry.

The null results at the LHC set stringent constraints on the lower bound of the masses of heavy gauge bosons which are assumed to be singly produced. 
Heavy $Z'$ from the aforementioned scenario can be pair-produced at the colliders, and currently the LHC puts loose limits. It was therefore proposed in this work to search for $2h +E_T^{miss}$ signatures at the LHC. The Littest Higgs Model with $T$-partiy was explored here as a  framework that realises the particle spectrum required for such signature. The $T$-odd partner of the SM $Z$-boson, $Z'$, decays exclusively to a Higgs boson and a dark matter $A'$, which is the massive $T$-odd partner of the photon.  We exploit (i) the large $E_T^{miss}$ signature due to $A'$, and (ii) boosted topology of the resulting Higgs due to the large $Z'$-$A'$ mass gap to discern the signal from the SM background events. Assuming $100\%$ decay branching of $Z' \to h A'$, the HL-LHC with $\mathcal{O}(100)~{\rm fb}^{-1}$ luminosity is able to discover the $Z'$ of mass around $500$ GeV.  And such a pair-produced $Z'$ is expected to yield important insights in electroweak symmetry breaking in relation to the dark sector. 


\acknowledgments
This work was partially supported the Ministry of Science and Technology of Taiwan under Grant No MOST-105-2112-M-003-010-MY3.  

\appendix

\begin{landscape}

\section{Appendix}
\vfill
\begin{centering}
\begin{table}[h]
\fontsize{10pt}{12pt}\selectfont
\begin{center}

{\renewcommand{\arraystretch}{1.25}\newcolumntype{C}[1]{>{\centering\let\newline\\\arraybackslash\hspace{0pt}}m{#1}} 
\begin{tabular}{ccc|cc|c|c|c}
\hline 
\multicolumn{3}{c|}{\multirow{2}{*}{Cuts}} & \multicolumn{5}{c}{   \hspace{8mm} $\sigma(pp\rightarrow Z' Z')$ [$10^{-4}$ fb]  \hspace{8mm} $\left(m_{Z_h}=5 \ m_{A_h}\right)$ } \\
\cline{4-8}
& & & \multicolumn{2}{c|}{ $m_{Z_h}=500 \ \text{GeV}$ } 
& \multirow{1}{*}{ $1000 \ \text{GeV}$ } & \multirow{1}{*}{ $1500 \ \text{GeV}$} &\multirow{1}{*}{ $2000 \ \text{GeV}$ } \\

\hline
(C0) & \multicolumn{2}{c|}{Basic cut} &  \multicolumn{2}{c|}{9.70 $\times 10^{3}$} & 2.73 $\times 10^{2}$ & 2.19 $\times 10^{1}$ & 2.45 \\
\hline
(C1) & \multicolumn{2}{c|}{ $N_\gamma \ge 2$ } & \multicolumn{2}{c|}{6.10 $\times 10^{3}$} & 1.64 $\times 10^{2}$ & 1.22 $\times 10^{1}$ & 1.17 \\
\hline
(C2) & \multicolumn{2}{c|}{ $N_\text{fatjet} \ge 1$ } & \multicolumn{2}{c|}{6.01 $\times 10^{3}$} & 1.63 $\times 10^{2}$ & 1.21 $\times 10^{1}$ & 1.16 \\
\hline
(C3) & \multicolumn{2}{c|}{ $N_\ell\leq  0$ } & \multicolumn{2}{c|}{6.00 $\times 10^{3}$} & 1.62 $\times 10^{2}$ & 1.21 $\times 10^{1}$ & 1.16 \\
\hline
(C4) & \multicolumn{2}{c|}{ $E_T^{miss} \geq 150~\text{GeV}$ } & \multicolumn{2}{c|}{4.91 $\times 10^{3}$} & 1.50 $\times 10^{2}$ & 1.16 $\times 10^{1}$ & 1.13 \\
\hline
(C5) & \multicolumn{2}{c|}{  $|m_{\gamma\gamma} -125~\text{GeV}|< 10 \ \text{GeV}$ } & \multicolumn{2}{c|}{4.84 $\times 10^{3}$} & 1.48 $\times 10^{2}$ & 1.14 $\times 10^{1}$ & 1.11 \\
\hline
(C6) & \multicolumn{2}{c|}{ $|m_{\text{fatjet}}-125~\text{GeV}|< 30 \ \text{GeV}$ } & \multicolumn{2}{c|}{3.23 $\times 10^{3}$} & 9.55 $\times 10^{1}$ & 7.27 & 6.75 $\times 10^{-1}$ \\
\hline
(C7) & \multicolumn{2}{c|}{ (1+)$b$ in fatjet } & \multicolumn{2}{c|}{ 1.24 $\times 10^{3}$ } & 3.66 $\times 10^{1}$ & 2.85 & 2.91 $\times 10^{-1}$  \\
\hline
(C8) & \multicolumn{2}{c|}{ $m_{hh} > 252\ \mathrm{GeV}$ } & \multicolumn{2}{c|}{1.22 $\times 10^{3}$ } & 3.59 $\times 10^{1}$ & 2.80 & 2.873 $\times 10^{-1}$ \\
\hline
(C9) & \multicolumn{2}{c|}{ $p_T^\text{fatjet} >200~\mathrm{GeV}$ } & \multicolumn{2}{c|}{1.20 $\times 10^{3}$ } & 3.56 $\times 10^{1}$ & 2.79 & 2.867 $\times 10^{-1}$ \\
\hline\hline
\end{tabular}}

\caption{Select cuts and resulting cross sections for the signals of different masses  14 TeV LHC. Couplings currently $g_{Z' A' h} = g_{qq'Z'}=0.6$.}

\end{center}
\end{table}
\end{centering}
\vfill
\end{landscape}

\bibliographystyle{mybibsty}
\bibliography{myrefs}

\end{document}